\documentstyle[twocolumn,aps,floats,amssymb,epsfig]{revtex}

\begin{document}

\def\d{{\rm d}}
\def\e{{\rm e}}
\def\O{{\rm O}}
\def\half{\mbox{$\frac12$}}
\def\eref#1{(\protect\ref{#1})}
\def\etal{{\it{}et~al.}}
\def\Li{\mathop{\rm Li}}
\def\av#1{\left\langle#1\right\rangle}
\def\set#1{\left\lbrace#1\right\rbrace}
\def\from{\leftarrow}

\draft
\tolerance = 10000

\renewcommand{\topfraction}{0.9}
\renewcommand{\textfraction}{0.1}
\renewcommand{\floatpagefraction}{0.9}
\setlength{\tabcolsep}{4pt}
\setcounter{topnumber}{1}

\twocolumn[\hsize\textwidth\columnwidth\hsize\csname @twocolumnfalse\endcsname

\title{Who is the best connected scientist?\\
A study of scientific coauthorship networks}
\author{M. E. J. Newman}
\address{Santa Fe Institute, 1399 Hyde Park Road, Santa Fe, NM 87501}
\address{Center for Applied Mathematics, Cornell University, Rhodes Hall,
Ithaca, NY 14853}
\maketitle

\begin{abstract}
Using data from computer databases of scientific papers in physics,
biomedical research, and computer science, we have constructed networks of
collaboration between scientists in each of these disciplines.  In these
networks two scientists are considered connected if they have coauthored
one or more papers together.  We have studied many statistical properties
of our networks, including numbers of papers written by authors, numbers of
authors per paper, numbers of collaborators that scientists have, typical
distance through the network from one scientist to another, and a variety
of measures of connectedness within a network, such as closeness and
betweenness.  We further argue that simple networks such as these cannot
capture the variation in the strength of collaborative ties and propose a
measure of this strength based on the number of papers coauthored by pairs
of scientists, and the number of other scientists with whom they coauthored
those papers.  Using a selection of our results, we suggest a variety of
possible ways to answer the question ``Who is the best connected
scientist?''
\end{abstract}


]

\vspace{1cm}

\section{Introduction}
A social network~\cite{WF94,Scott00} is a set of people or groups each of
which has connections of some kind to some or all of the others.  In the
language of social network analysis, the people or groups are called
``actors'' and the connections ``ties.''  Both actors and ties can be
defined in different ways depending on the questions of interest.  An actor
might be a single person, a team, or a company.  A tie might be a
friendship between two people, a collaboration or common member between two
teams, or a business relationship between companies.

Social network analysis has a history stretching back at least half a
century, and has produced many results concerning social influence, social
groupings, inequality, disease propagation, communication of information,
and indeed almost every topic that has interested twentieth century
sociology.  This paper, however, is a physics paper.  Why should a
physicist be interested in social networks?  There has, in fact, been a
substantial surge of interest in social networks within the physics
community recently, as evidenced by the large body of papers on the
topic---see Refs.~\onlinecite{Review00} through~\onlinecite{NSW00} and
references therein.  The techniques of statistical physics in particular
turn out to be well suited to the study of these networks.  Profitable use
has been made of a variety of physical modeling
techniques~\cite{WS98,Barabasi99,Kumar00}, exact
solutions~\cite{Moukarzel99,DM00,KRL00,DMS00,KAS00,Kleinberg00}, Monte
Carlo simulation~\cite{AJB99,BA99,NW99,MMP00}, scaling and renormalization
group methods~\cite{BA99,NW99,MMP00}, mean-field theory~\cite{BAJ99,NMW00},
percolation theory~\cite{MN00,CEBH00,Callaway00}, the replica
method~\cite{BW00}, generating functions~\cite{MN00,Callaway00,NSW00}, and
a host of other techniques familiar to physicists.

This paper makes use of some of these techniques in the study of some
specific examples of social networks.  However, the subject matter of this
paper will be of interest to physicists for another reason: it's about
them.  In this paper, we study collaboration networks in which the actors
are scientists, and the ties between them are scientific collaborations, as
documented in the papers that they write.

\section{Collaboration networks}
\label{collabnets}
Traditional investigations of social networks have been carried out through
field studies.  Typically one looks at a fairly self-contained community
such as a business community~\cite{Mariolis75,GM78,PA93}, a
school~\cite{FRO63,FS64}, a religious or ethnic community~\cite{Bernard88},
and so forth, and constructs the network of ties by interviewing
participants, or by circulating questionnaires.  A study will ask
respondents to name those with whom they have the closest ties, probably
ranked by subjective closeness, and may optionally call for additional
information about those people or about the nature of the ties.

Studies of this kind have revealed much about the structure of communities,
but they suffer from two substantial problems which make them poor sources
of data for the kinds of quantitative approach to network analysis that
physics has adopted.  First, the data they return are not numerous.
Collecting and compiling data from these studies is an arduous process and
most data sets contain no more than a few tens or hundreds of actors.  It
is a rare study that exceeds a thousand actors.  This makes the statistical
accuracy of many results poor, a particular difficulty for the
large-system-size methods adopted by statistical physics.  Second, they
contain significant and uncontrolled errors as a result of the subjective
nature of the respondents' replies.  What one respondent considers to be a
friendship or acquaintance, for example, may be completely different from
what another respondent does.  In studies of
school-children~\cite{FRO63,FS64}, for instance, it is found that some
children will claim friendship with every single one of their hundreds of
schoolmates, while others will name only one or two friends.  Clearly these
respondents are employing different definitions of friendship.

In response to these inadequacies, many researchers have turned instead to
other, better documented networks, for which reliable statics can be
collected.  Examples include the world-wide
web~\cite{AJB99,Broder00,AJB00}, power grids~\cite{WS98}, telephone call
graphs~\cite{ABW98}, and airline timetables~\cite{ASBS00}.  These graphs
are certainly interesting in their own right, and furthermore might loosely
be regarded as social networks, since their structure clearly reflects
something about the structure of the society that built them.  However,
their connection to the ``true'' social networks discussed here is tenuous
at best and so, for our purposes, they cannot offer a great deal of
insight.

A more promising source of data is the affiliation network.  An affiliation
network is a network of actors connected by common membership in groups of
some sort, such as clubs, teams, or organizations.  Examples which have
been studied in the past include women and the social events they
attend~\cite{DGG41}, company CEOs and the clubs they frequent~\cite{GM78},
company directors and the boards of directors on which they
sit~\cite{Mariolis75,DG97}, and movie actors and the movies in which they
appear~\cite{WS98,ASBS00}.  Data on affiliation networks tend to be more
reliable than those on other social networks, since membership of a group
can often be determined with a precision not available when considering
friendship or other types of acquaintance.  Very large networks can be
assembled in this way as well, since in many cases group membership can be
ascertained from membership lists, making time-consuming interviews or
questionnaires unnecessary.  A network of movie actors, for example, has
been compiled using the resources of the Internet Movie
Database~\cite{IMDB}, and contains the names of nearly half a million
actors---a much better sample on which to perform statistics than most
social networks, although it is unclear whether this particular network has
any real social interest.

In this paper we construct affiliation networks of scientists in which a
link between two scientists is established by their coauthorship of one or
more scientific papers.  Thus the groups to which scientists belong in this
network are the groups of coauthors of a single paper.  This network is in
some ways more truly a social network than many affiliation networks; it is
probably fair to say that most pairs of people who have written a paper
together are genuinely acquainted with one another, in a way that movie
actors who appeared together in a movie may not be.  There are
exceptions---some very large collaborations, for example in high-energy
physics, will contain coauthors who have never even met---and we will
discuss these at the appropriate point.  By and large, however, the network
reflects genuine professional interaction between scientists, and may be
the largest social network ever studied~\cite{note1}.

The idea of constructing a network of coauthorship is not new.  Many
readers will be familiar with the concept of the Erd\"os number, named
after Paul Erd\"os, the Hungarian mathematician, one of the founding
fathers of graph theory amongst other things~\cite{Hoffman98}.  At some
point, it became a popular cocktail party pursuit for mathematicians to
calculate how far removed they were in terms of publication from Erd\"os.
Those who had published a paper with Erd\"os were given a Erd\"os number
of~1, those who had published with one of those people but not with
Erd\"os, a number of~2, and so forth.  In the jargon of social networks,
your Erd\"os number is the geodesic distance between you and Erd\"os in the
coauthorship network.  In recent studies~\cite{GI95,BM00}, it has been
found that the average Erd\"os number is about 4.7, and the maximum known
finite Erd\"os number (within mathematics) is~15.  These results are
probably influenced to some extent by Erd\"os' prodigious mathematical
output: he published at least 1401 papers, more than any other
mathematician ever except possibly Leonhard Euler.  However, quantitatively
similar, if not quite so impressive, results are in most cases found if the
network is centered on other mathematicians.  (On the other hand,
fifth-most published mathematician, Lucien Godeaux, produced 644 papers, on
643 of which he was the sole author.  He has no finite Erd\"os
number~\cite{GI95}.  Clearly sheer size of output is not a sufficient
condition for high connectedness.)

There is also a substantial body of work in bibliometrics (a specialty
within information science) on extraction of collaboration patterns from
publication data---see Refs.~\onlinecite{ER90,PB95,MP96,Kretschmer98,DFC99}
and references therein.  However, these studies have not so far attempted
to reconstruct actual collaboration networks from bibliographic data,
concentrating more on organization and institutional aspects of
collaboration~\cite{note2}.

In this paper, we study networks of scientists using bibliographic data
drawn from four publicly available databases of papers.  The databases are:
\begin{enumerate}
\item Los Alamos e-Print Archive: a database of unrefereed preprints in
physics, self-submitted by their authors, running from 1992 to the present.
This database is subdivided into specialties within physics, such as
condensed matter and high-energy physics.
\item MEDLINE: a database of articles on biomedical research published in
refereed journals, stretching from 1961 to the present.  Entries in the
database are updated by the database's maintainers, rather than papers'
authors, giving it relatively thorough coverage of its subject area.  The
inclusion of biomedicine is crucial in a study such as this one.  In most
countries biomedical research easily dwarfs civilian research on any other
topic, in terms of both expenditure and human effort.  Any study which
omitted it would be leaving out the largest part of current scientific
research.
\item SPIRES: a database of preprints and published papers in high-energy
physics, both theoretical and experimental, from 1974 to the present.  The
contents of this database are also professionally maintained.  High energy
physics is an interesting case socially, having a tradition of much larger
experimental collaborations than any other discipline.
\item NCSTRL: a database of preprints in computer science, submitted by
participating institutions and stretching back about ten years.
\end{enumerate}

We have constructed networks of collaboration for each of these databases
separately, and analyzed them using a variety of techniques, some standard,
some invented for the purpose.  The outline of this paper is as follows.
In Section~\ref{basic} we discuss some basic statistics, to give a feel for
the shape of our networks.  Among other things we discuss the typical
numbers of authors, numbers of papers per author, authors per paper, and
number of collaborators of scientists in the various disciplines.  In
Section~\ref{centrality} we look at a variety of measures concerned with
paths between scientists on the network.  In Section~\ref{weighted} we
extend our networks to include a measure of the strength of collaborative
ties between scientists and examine measures of connectedness in these
weighted networks.  In Section~\ref{concs} we give our conclusions.  A
brief report of some of the work described here has appeared previously as
Ref.~\onlinecite{Newman00}.

\section{Basic results}
\label{basic}
For this study, we constructed collaboration networks using data from a
five year period from 1995 to 1999 inclusive, although data for much longer
periods were available in some of the databases.  There were several
reasons for using this fairly short time window.  First, older data are
less complete than newer for all databases.  Second, we wanted to study the
same time period for all databases, so as to be able to make valid
comparisons between collaboration patterns in different fields.  The
coverage provided by both the Los Alamos Archive and the NCSTRL database is
relatively poor before 1995, and this sets a limit on how far back we can
look.  Third, networks change over time, both because people enter and
leave the professions they represent and because practices of scientific
collaboration and publishing change.  In this particular study we have not
examined time evolution in the network, although this is certainly an
interesting topic for research and indeed is currently under investigation
by others~\cite{Barabasi00}.  For our purposes, a short window of data is
desirable, to ensure that the collaboration network is roughly static
during the study.

The raw data for the networks described here are computer files containing
lists of papers, including authors names and possibly other information
such as title, abstract, date, journal reference, and so forth.
Construction of the collaboration networks is straightforward.  The files
are parsed to extract author names and as names are found a list is
maintained of the ones seen so far---vertices already in the network---so
that recurring names can be correctly assigned to extant vertices.  Edges
are added between each pair of authors on each paper.  A naive
implementation of this calculation, in which names are stored in a simple
array, would take time $\O(pn)$, where $p$ is the total number of papers in
the database and $n$ the number of authors.  This however turns out to be
prohibitively slow for large networks since $p$ and $n$ are of similar size
and may be a million or more.  Instead therefore, we store the names of the
authors in an ordered binary tree, which reduces the running time to
$\O(p\log n)$, making the calculation tractable, even for the largest
databases studied here.

\begin{table*}
\setlength{\tabcolsep}{8.5pt}
\begin{center}
\begin{tabular}{l|c|cccc|c|c}
 &         & \multicolumn{4}{c|}{Los Alamos e-Print Archive}           &        \\
\cline{3-6}
 & MEDLINE & complete & {\tt astro-ph} & {\tt cond-mat} & {\tt hep-th} & SPIRES & NCSTRL \\
\hline
 total papers               & $2163923$    & $98502$     & $22029$     & $22016$     & $19085$     & $66652$       & $13169$       \\
 total authors              & $1520251$    & $52909$     & $16706$     & $16726$     & $8361$      & $56627$       & $11994$       \\
 \quad first initial only   & $1090584$    & $45685$     & $14303$     & $15451$     & $7676$      & $47445$       & $10998$       \\
 mean papers per author     & $6.4(6)$     & $5.1(2)$    & $4.8(2)$    & $3.65(7)$   & $4.8(1)$    & $11.6(5)$     & $2.55(5)$     \\
 mean authors per paper     & $3.754(2)$   & $2.530(7)$  & $3.35(2)$   & $2.66(1)$   & $1.99(1)$   & $8.96(18)$    & $2.22(1)$     \\
 collaborators per author   & $18.1(1.3)$  & $9.7(2)$    & $15.1(3)$   & $5.86(9)$   & $3.87(5)$   & $173(6)$      & $3.59(5)$     \\
 size of giant component    & $1395693$    & $44337$     & $14845$     & $13861$     & $5835$      & $49002$       & $6396$        \\
 \quad first initial only   & $1019418$    & $39709$     & $12874$     & $13324$     & $5593$      & $43089$       & $6706$        \\
 \quad as a percentage      & $92.6(4)\%$  & $85.4(8)\%$ & $89.4(3)$   & $84.6(8)\%$ & $71.4(8)\%$ & $88.7(1.1)\%$ & $57.2(1.9)\%$ \\
 2nd largest component      & $49$         & $18$        & $19$        & $16$        & $24$        & $69$          & $42$          \\
 clustering coefficient     & $0.066(7)$   & $0.43(1)$   & $0.414(6)$  & $0.348(6)$  & $0.327(2)$  & $0.726(8)$    & $0.496(6)$    \\
 mean distance              & $4.6(2)$     & $5.9(2)$    & $4.66(7)$   & $6.4(1)$    & $6.91(6)$   & $4.0(1)$      & $9.7(4)$      \\
 maximum distance           & $24$         & $20$        & $14$        & $18$        & $19$        & $19$          & $31$          \\
\end{tabular}
\end{center}
\caption{Summary of results of the analysis of seven scientific
  collaboration networks.  Numbers in parentheses are standard errors
  on the least significant figures.}
\label{summary}
\end{table*}

In Table~\ref{summary} we give a summary of some of the basic results for
the networks studied here.  We discuss these results in detail in the rest
of this section.

\subsection{Number of authors}
The size of the databases varies considerably from about a million authors
for MEDLINE to about ten thousand for NCSTRL.  In fact, it is difficult to
say with precision how many authors there are.  One can say how many
distinct {\em names\/} appear in a database, but the number of names is not
the same as the number of authors.  A single author may report their name
differently on different papers.  For example, F. L. Wright, Francis
Wright, and Frank Lloyd Wright could all be the same person.  Also two
authors may have the same name.  Grossman and Ion~\cite{GI95} point out
that there are two American mathematicians named Norman Lloyd Johnson, who
are known to be distinct people and who work in different fields, but
between whom computer programs such as ours cannot hope to distinguish.
Even additional clues such as home institution or field of specialization
cannot be used to distinguish such people, since many scientists have more
than one institution or publish in more than one field.  The present
author, for example, has addresses at the Santa Fe Institute and Cornell
University, and publishes in both statistical physics and paleontology.

In order to control for these biases, we constructed two different versions
of each of the collaboration networks studied here, as follows.  In the
first, we identify each author by his or her surname and first initial
only.  This method is clearly prone to confusing two people for one, but
will rarely fail to identify two names which genuinely refer to the same
person.  In the second version of each network, we identify authors by
surname and all initials.  This method can much more reliably distinguish
authors from one another, but will also identify one person as two if they
give their initials differently on different papers.  Indeed this second
measure appears to overestimate the number of authors in a database
substantially.  Networks constructed in these two different fashions
therefore give upper and lower bounds on the number of authors, and hence
also give bounds on many of the other quantities studied here.  In
Table~\ref{summary} we give numbers of authors in each network using both
methods, but for many of the other quantities we give only an error
estimate based on the separation of the bounds.

\subsection{Number of papers per author}
The average number of papers per author in the various subject areas is in
the range of around three to six over the five year period.  The only
exception is the SPIRES database, covering high-energy physics, in which
the figure is significantly higher at $11.6$.  One possible explanation for
this is that SPIRES is the only database which contains both preprints and
published papers.  It is possible that the high figure for papers per
author reflects duplication of papers in both preprint and published form.
However, the maintainers of the database go to some lengths to avoid
this~\cite{OConnell00}, and a more probable explanation is perhaps that
publication rates are higher for the large collaborations favored by
high-energy physics, since a large group of scientists has more
person-hours available for the writing of papers.

\begin{figure}
\begin{center}
\psfig{figure=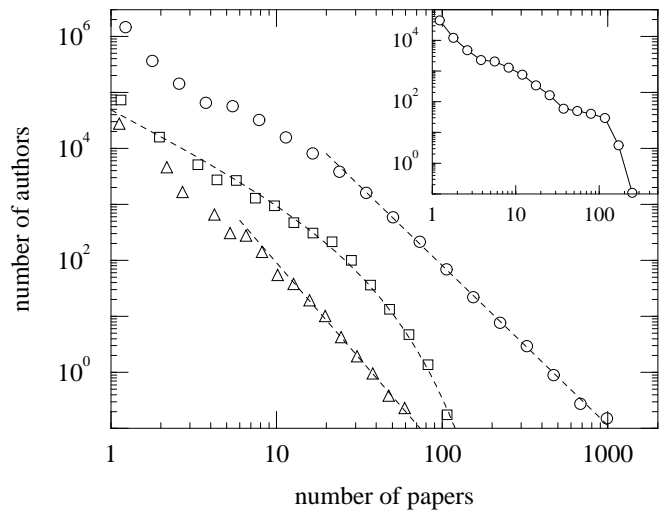,width=\columnwidth}
\end{center}
\caption{Histograms of the number of papers written by authors in MEDLINE
(circles), the Los Alamos Archive (squares), and NCSTRL (triangles).  The
dotted lines are fits to the data as described in the text.  Inset: the
equivalent histogram for the SPIRES database.}
\label{papers}
\end{figure}

\begin{table*}[t]
\begin{tabular}{c|rl|rl|rl|rl}
 & \multicolumn{2}{c|}{number of papers} & \multicolumn{2}{c|}{number of co-workers} & \multicolumn{2}{c|}{betweenness ($\times10^6$)} &
 \multicolumn{2}{c}{collaboration weight} \\
\hline
\psfig{file=label-ap.eps,width=10pt,rheight=0pt}
 & 112 & Fabian, A.C.      & 360 & Frontera, F.     & 2.33 & Kouveliotou, C.  & 16.5 & Moskalenko, I.V./Strong, A.W.   \\
 & 101 & van Paradijs, J.  & 353 & Kouveliotou, C.  & 2.15 & van Paradijs, J. & 15.0 & Hernquist, L./Heyl, J.S.        \\
 &  81 & Frontera, F.      & 329 & van Paradijs, J. & 1.80 & Filippenko, A.V. & 14.0 & Mathews, W.G./Brighenti, F.     \\
 &  80 & Hernquist, L.     & 299 & Piro, L.         & 1.57 & Beaulieu, J.P.   & 13.4 & Labini, F.S./Pietronero, L.     \\
 &  79 & Gould, A.         & 296 & Costa, E.        & 1.52 & Nomoto, K.       & 12.2 & Piran, T./Sari, R.              \\
 &  78 & Silk, J.          & 291 & Feroci, M.       & 1.52 & Pian, E.         & 11.8 & Zaldarriaga, M./Seljak, U.      \\
 &  78 & Klis, M.V.D.      & 284 & Pian, E.         & 1.49 & Frontera, F.     & 11.4 & Hernquist, L./Katz, N.          \\
 &  73 & Kouveliotou, C.   & 284 & Hurley, K.       & 1.35 & Silk, J.         & 11.1 & Avila-Reese, V./Firmani, C.     \\
 &  70 & Ghisellini, G.    & 244 & Palazzi, E.      & 1.33 & Kamionkowski, M. & 10.9 & Dai, Z.G./Lu, T.                \\
 &  66 & Piro, L.          & 244 & Heise, J.        & 1.28 & McMahon, R.G.    & 10.8 & Ostriker, J.P./Cen, R.          \\
\hline
\psfig{file=label-cm.eps,width=10pt,rheight=0pt}
 & 116 & Parisi, G.        & 107 & Uchida, S.       & 4.11 & MacDonald, A.H.  & 22.3 & Belitz, D./Kirkpatrick, T.R.    \\
 &  79 & Scheffler, M.     & 103 & Ueda, Y.         & 3.96 & Bishop, A.R.     & 17.0 & Shrock, R./Tsai, S.             \\
 &  75 & Das Sarma, S.     &  96 & Revcolevschi, A. & 3.36 & Das Sarma, S.    & 15.0 & Yukalov, V.I./Yukalova, E.P.    \\
 &  74 & Stanley, H.E.     &  94 & Eisaki, H.       & 2.96 & Tosatti, E.      & 14.7 & Mart\'\i{}n-Delgado, M.A./Sierra, G. \\
 &  70 & MacDonald, A.H.   &  84 & Cheong, S.       & 2.52 & Wang, X.         & 14.3 & Krapivsky, P.L./Ben-Naim, E.    \\
 &  68 & Sornette, D.      &  83 & Isobe, M.        & 2.38 & Revcolevschi, A. & 14.1 & Beenakker, C.W.J./Brouwer, P.W. \\
 &  60 & Volovik, G.E.     &  78 & Stanley, H.E.    & 2.30 & Uchida, S.       & 13.8 & Weng, Z.Y./Sheng, D.N.          \\
 &  56 & Beenakker, C.W.J. &  76 & Shirane, G.      & 2.21 & Sigrist, M.      & 13.7 & Sornette, D./Johansen, A.       \\
 &  53 & Dagotto, E.       &  76 & Scheffler, M.    & 2.19 & Cheong, S.       & 13.6 & Rikvold, P.A./Novotny, M.A.     \\
 &  50 & Helbing, D.       &  76 & Menovsky, A.A.   & 2.18 & Stanley, H.E.    & 13.0 & Scalapino, D.J./White, S.R.     \\
\hline
\psfig{file=label-ht.eps,width=10pt,rheight=0pt}
 & 78 & Odintsov, S.D.     & 50 & Ambjorn, J.       & 0.98 & Odintsov, S.D.   & 34.0 & Lu, H./Pope, C.N.               \\
 & 73 & Lu, H.             & 44 & Ferrara, S.       & 0.88 & Ambjorn, J.      & 29.0 & Odintsov, S.D./Nojiri, S.       \\
 & 72 & Pope, C.N.         & 43 & Vafa, C.          & 0.88 & Kogan, I.I.      & 18.7 & Lee, H.W./Myung, Y.S.           \\
 & 69 & Cvetic, M.         & 39 & Odintsov, S.D.    & 0.84 & Henneaux, M.     & 18.3 & Schweigert, C./Fuchs, J.        \\
 & 68 & Ferrara, S.        & 39 & Kogan, I.I.       & 0.73 & Douglas, M.R.    & 14.7 & Ovrut, B.A./Waldram, D.         \\
 & 65 & Vafa, C.           & 36 & Proeyen, A.V.     & 0.67 & Ferrara, S.      & 14.7 & Kleihaus, B./Kunz, J.           \\
 & 65 & Tseytlin, A.A.     & 35 & Fre, P.           & 0.63 & Vafa, C.         & 12.9 & Mavromatos, N.E./Ellis, J.      \\
 & 65 & Mavromatos, N.E.   & 35 & Ellis, J.         & 0.60 & Khare, A.        & 12.4 & Kachru, S./Silverstein, E.      \\
 & 63 & Witten, E.         & 35 & Douglas, M.R.     & 0.58 & Tseytlin, A.A.   & 11.7 & Kakushadze, Z./Tye, S.H.H.      \\
 & 54 & Townsend, P.K.     & 34 & Lu, H.            & 0.58 & Townsend, P.K.   & 11.6 & Arefeva, I.Y./Volovich, I.V.    \\
\end{tabular}
\caption{The authors with the highest numbers of papers, numbers of
coauthors, and betweenness, and strongest collaborations in astrophysics,
condensed matter physics, and high-energy theory.  The figures for
betweenness have been divided by $10^6$.  Full lists of the rankings of all
the authors in these databases can be found on the world-wide
web~\protect\cite{note3}.}
\label{top10}
\end{table*}

In addition to the average numbers of papers per author in each database,
it is interesting to look at the distribution $p_k$ of numbers $k$ of
papers per author.  In 1926, Alfred Lotka~\cite{Lotka26} showed, using a
dataset compiled by hand, that this distribution followed a power law, with
exponent approximately~$-2$, a result which is now referred to as Lotka's
Law of Scientific Productivity.  In other words, in addition to the many
authors who publish only a small number of papers, one expects to see a
``fat tail'' consisting of a small number of authors who publish a very
large number of papers.  In Fig.~\ref{papers} we show on logarithmic scales
histograms for each of our four databases of the numbers of papers
published.  (These histograms and all the others shown here were created
using the ``all initials'' versions of the collaboration networks.)  For
the MEDLINE and NCSTRL databases these histograms follow a power law quite
closely, at least in their tails, with exponents of $-2.86(3)$ and
$-3.41(7)$ respectively---somewhat steeper than those found by Lotka, but
in reasonable agreement with other more recent
studies~\cite{ER90,Voos74,Pao86}.  For the Los Alamos Archive the pure
power law is a poor fit.  An exponentially truncated power law does much
better:
\begin{equation}
p_k = C k^{-\tau} \e^{-k/\kappa},
\label{cutoff}
\end{equation}
where $\tau$ and $\kappa$ are constants and $C$ is fixed by the requirement
of normalization---see Fig.~\ref{papers}.  (The probability $p_0$ of having
zero papers is taken to be zero, since the names of scientists who have not
written any papers do not appear in the database.)  The exponential cutoff
we attribute to the finite time window of five years used in this study
which prevents any one author from publishing a very large number of
papers.  Lotka and subsequent authors who have confirmed his law have not
usually used such a window.

It is interesting to speculate why the cutoff appears only in physics and
not in computer science or biomedicine.  Surely the five year window limits
everyone's ability to publish very large numbers of papers, regardless of
their area of specialization?  For the case of MEDLINE one possible
explanation is suggested by a brief inspection of the names of the most
published authors.  It turns out that most of the names of highly published
authors in MEDLINE are Japanese (with a sprinkling of Chinese names among
them).  The top ten, for example, are Suzuki,~T., Wang,~Y., Suzuki,~K.,
Takahashi,~M., Nakamura,~T., Tanaka,~K., Tanaka,~T., Wang,~J., Suzuki,~Y.,
and Takahashi,~T\null.  This may reflect differences in author attribution
practices in Japanese biomedical research, but more probably these are
simply common names, and these apparently highly published authors are each
several different people who have been conflated in our analysis.  Thus it
is possible that there is not after all any fat tail in the distribution
for the MEDLINE database, only the illusion of one produced by the large
number of scientists with commonly occurring names.  (This doesn't however
explain why the tail appears to follow a power law.)  This argument is
strengthened by the sheer numbers of papers involved.  T.~Suzuki published,
it appears, 1697 papers, or about one paper a day, including weekends and
holidays, every day for the entire course of our five year study.  This
seems to be an improbably large output.

Interestingly, no national bias is seen in any of the other databases, and
the names which top the list in physics and computer science are not common
ones.  (For example, the most published authors in the other three
databases are Shelah,~S. (Los Alamos Archive), Wolf,~G. (SPIRES), and
Bestavros,~A. (NCSTRL).)  Thus it is still unclear why the NCSTRL database
should have a power-law tail, though this database is small and it is
possible that it does possess a cutoff in the productivity distribution
which is just not visible because of the limits of the dataset.

For the SPIRES database, which is shown separately in the inset of the
figure, neither pure nor truncated power law fits the data well, the
histogram displaying a significant bump around the 100-paper mark.  A
possible explanation for this is that a small number of large
collaborations published around this number of papers during the
time-period studied.  Since each author in such a collaboration is then
credited with publishing a hundred papers, the statistics in the tail of
the distribution can be substantially skewed by such practices.

In the first column of Table~\ref{top10}, we list the most frequent authors
in three subject-specific subdivisions of the Los Alamos Archive: {\tt
astro-ph} (astro-physics), {\tt cond-mat} (condensed matter physics), and
{\tt hep-th} (high-energy theory).
Although there is only space to list the top ten winners in this table, the
entire list (and the corresponding lists for the other tables in this
paper) can be found by the curious reader on the world-wide
web~\cite{note3}.

\subsection{Numbers of authors per paper}
Grossman and Ion~\cite{GI95} report that the average number of authors on
papers in mathematics has increased steadily over the last sixty years,
from a little over 1 to its current value of about~$1.5$.  Higher numbers
still seem to apply to current studies in the sciences.  Purely theoretical
papers appear to be typically the work of two scientists, with high-energy
theory and computer science showing averages of $1.99$ and $2.22$ in our
calculations.  For databases covering experimental or partly experimental
subject areas the averages are, not surprisingly, higher: $3.75$ for
biomedicine, $3.35$ for astrophysics, $2.66$ for condensed matter physics.
The SPIRES high-energy physics database however shows the most startling
results, with an average of $8.96$ authors per paper, obviously a result of
the presence of papers in the database written by very large
collaborations.  (Perhaps what is most surprising about this result is
actually how small it is.  The hundreds strong mega-collaborations of CERN
and Fermilab are sufficiently diluted by theoretical and smaller
experimental groups, that the number is only 9, and not 100.)

\begin{figure}
\begin{center}
\psfig{figure=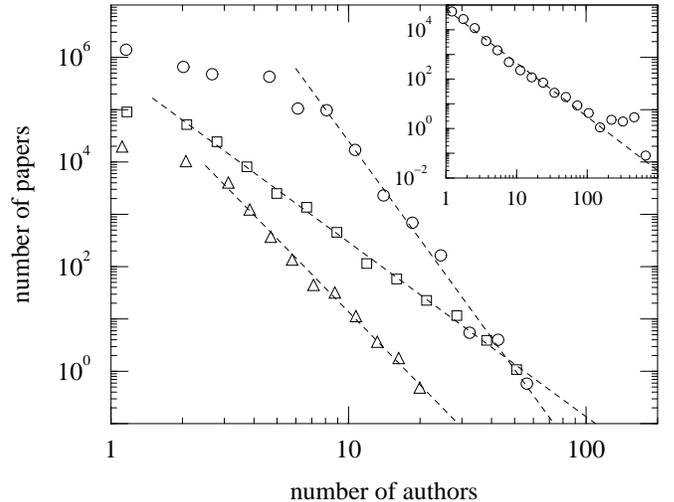,width=\columnwidth}
\end{center}
\caption{Histograms of the number of authors on papers in MEDLINE
(circles), the Los Alamos Archive (squares), and NCSTRL (triangles).  The
dotted lines are the best fit power-law forms.  Inset: the equivalent
histogram for the SPIRES database, showing a clear peak in the 200 to 500
author range.}
\label{authors}
\end{figure}

Distributions of numbers of authors per paper are shown in
Fig.~\ref{authors}, and appear to have power-law tails with widely varying
exponents of $-6.2(3)$ (MEDLINE), $-3.34(5)$ (Los Alamos Archive),
$-4.6(1)$ (NCSTRL), and $-2.18(7)$ (SPIRES).  The SPIRES data, which are
again shown in a separate inset, also display a pronounced peak in the
distribution around 200--500 authors.  This peak presumably corresponds to
the large experimental collaborations which dominate the upper end of this
histogram.

The largest number of authors on a single paper was 1681 (in high-energy
physics, of course).

\subsection{Numbers of collaborators per author}
The differences between the various disciplines represented in the
databases are emphasized still more by the numbers of collaborators that a
scientist has, the total number of people with whom a scientist wrote
papers during the five-year period.  The average number of collaborators is
markedly lower in the purely theoretical disciplines ($3.87$ in high-energy
theory, $3.59$ in computer science) than in the wholly or partly
experimental ones ($18.1$ in biomedicine, $15.1$ in astrophysics).  But the
SPIRES high-energy physics database takes the prize once again, with
scientists having an impressive 173 collaborators, on average, over a
five-year period.  This clearly begs the question whether the high-energy
coauthorship network can be considered an accurate representation of the
high-energy physics community at all; it seems unlikely that an author
could know 173 colleagues well.

\begin{figure}
\begin{center}
\psfig{figure=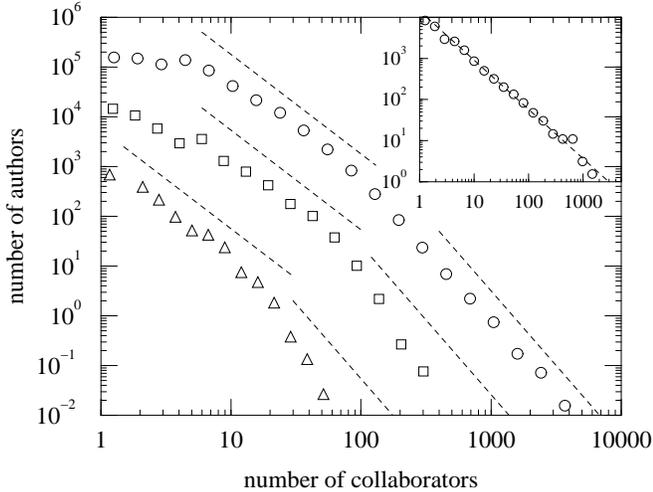,width=\columnwidth}
\end{center}
\caption{Histograms of the number of collaborators of authors in MEDLINE
(circles), the Los Alamos Archive (squares), and NCSTRL (triangles).  The
dotted lines show how power-law distributions with exponents $-2$ and $-3$
would look on the same axes.  Inset: the equivalent histogram for the
SPIRES database, which is well fit by a single power law (dotted line).}
\label{collabs}
\end{figure}

The distributions of numbers of collaborators are shown in
Fig.~\ref{collabs}.  In all cases they appear to have long tails, but only
the SPIRES data (inset) fit a power-law distribution well, with a low
measured exponent of~$-1.20$.  Note also the small peak in the SPIRES data
around 700---presumably again a result of the presence of large
collaborations.

For the other three databases, the distributions show some curvature.  This
may, as we have previously suggested~\cite{Newman00}, be the signature of
an exponential cutoff, produced once again by the finite time window of the
study.  (Redner~\cite{RednerPC} has suggested an alternative origin for the
cutoff using growth models of networks---see Ref.~\onlinecite{KRL00}.)  An
alternative possibility has been suggested by Barab\'asi~\cite{BarabasiPC},
based on models of the collaboration process.  In one such
model~\cite{Barabasi00}, the distribution of the number of collaborators of
an author follows a power law with slope $-2$ initially, changing to slope
$-3$ in the tail, the position of the crossover depending on the length of
time for which the collaboration network has been evolving.  We show slopes
$-2$ and $-3$ as dotted lines on the figure, and the agreement with the
curvature seen in the data is moderately good, particularly for the MEDLINE
data.  (For the Los Alamos and NCSTRL databases, the slope in the tail
seems to be somewhat steeper than~$-3$.)

Column 2 of Table~\ref{top10} shows the authors in {\tt astro-ph}, {\tt
cond-mat}, and {\tt hep-th} with the largest numbers of collaborators.  The
winners in this race tend to be experimentalists, who conduct research in
larger groups, though there are exceptions.  The high-energy theory
database of course contains only theorists, and the smaller numbers of
collaborators reflects this.

\subsection{Size of the giant component}
\label{gc}
In the theory of random graphs~\cite{NSW00,ER60,Bollobas85,MR9598} it is
known that there is a continuous phase transition with increasing density
of edges in a graph at which a ``giant component'' forms, i.e.,~a connected
subset of vertices whose size scales extensively.  Well above this
transition, in the region where the giant component exists, the giant
component usually fills a large portion of the graph, and all other
components (i.e.,~connected subsets of vertices) are small, typically of
size $\O(\log n)$, where $n$ is the total number of vertices.  We see a
situation reminiscent of this in all of the graphs studied here: a single
large component of connected vertices which fills the majority of the
volume of the graph, and a number of much smaller components filling the
rest.  In Table~\ref{summary} we show the size of the giant component for
each of our databases, both as total number of vertices and as a fraction
of system size.  In all cases the giant component fills around 80\% or 90\%
of the total volume, except for high-energy theory and computer science,
which give smaller figures.  A possible explanation of these two anomalies
may be that the corresponding databases give poorer coverage of their
subjects.  The {\tt hep-th} high-energy database is quite widely used in
the field, but overlaps to a large extent with the longer established
SPIRES database, and it is possible that some authors neglect it for this
reason~\cite{OConnell00}.  The NCSTRL computer science database differs
from the others in this study in that the preprints it contains are
submitted by participating institutions, of which there are about 160.
Preprints from institutions not participating are mostly left out of the
database, and its coverage of the subject area is, as a result, incomplete.

The figure of 80--90\% for the size of the giant component is a promising
one.  It indicates that the vast majority of scientists are connected via
collaboration, and hence via personal contact, with the rest of their
field.  Despite the prevalence of journal publishing and conferences in the
sciences, person-to-person contact is still of paramount importance in the
communication of scientific information, and it is reasonable to suppose
that the scientific enterprise would be significantly hindered if
scientists were not so well connected to one another.

\subsection{Clustering coefficients}
An interesting idea circulating in social network theory currently is that
of ``transitivity,'' which, along with its sibling ``structural balance,''
describes symmetry of interaction amongst trios of actors.
``Transitivity'' has a different meaning in sociology from its meaning in
mathematics and physics, although the two are related.  It refers to the
extent to which the existence of ties between actors A and B and between
actors B and C implies a tie between A and~C.  The transitivity, or more
precisely the fraction of transitive triples, is that fraction of connected
triples of vertices which also form ``triangles'' of interaction.  Here a
connected triple means an actor who is connected to two others.  In the
physics literature, this quantity is usually called the clustering
coefficient~$C$~\cite{WS98}, and can be written
\begin{equation}
C = {\mbox{$3\times$ number of triangles on the graph}\over
     \mbox{number of connected triples of vertices}}.
\label{defsc}
\end{equation}
The factor of three in the numerator compensates for the fact that each
complete triangle of three vertices contributes three connected triples,
one centered on each of the three vertices, and ensures that $C=1$ on a
completely connected graph.  On all unipartite random graphs
$C=\O(n^{-1})$~\cite{WS98,NSW00}, where $n$ is the number of vertices, and
hence goes to zero in the limit of large graph size.  In social networks it
is believed that the clustering coefficient will take a non-zero value even
in very large networks, because there is a finite (and probably quite
large) probability that two people will be acquainted if they have another
acquaintance in common.  This is a hypothesis we can test with our
collaboration networks.  In Table~\ref{summary} we show values of the
clustering coefficient $C$, calculated from Eq.~\eref{defsc}, for each of
the databases studied, and as we see, the values are indeed large, as large
as $0.7$ in the case of the SPIRES database, and around $0.3$ or $0.4$ for
most of the others.

There are a number of possible explanations for these high values of~$C$.
First of all, it may be that they indicate simply that collaborations of
three or more people are common in science.  Every paper which has three
authors clearly contributes a triangle to the numerator of Eq.~\eref{defsc}
and hence increases the clustering coefficient.  This is, in a sense, a
``trivial'' form of clustering, although it is by no means socially
uninteresting.

In fact it turns out that this effect can account for some but not all of
the clustering seen in our graphs.  One can construct a random graph model
of a collaboration network which mimics the trivial clustering effect, and
the results indicate that only about a half of the clustering which we see
is a result of authors collaborating in groups of three or
more~\cite{NSW00}.  The rest of the clustering must have a social
explanation, and there are some obvious possibilities:
\begin{enumerate}
\item A scientist may collaborate with two colleagues individually, who may
then become acquainted with one another through their common collaborator,
and so end up collaborating themselves.  This is the usual explanation for
transitivity in acquaintance networks~\cite{WF94}.
\item Three scientists may all revolve in the same circles---read the same
journals, attend the same conferences---and, as a result, independently
start up separate collaborations in pairs, and so contribute to the value
of $C$, although only the workings of the community, and not any specific
person, is responsible for introducing them.
\item As a special case of the previous possibility---and perhaps the most
likely case---three scientists may all work at the same institution, and as
a result may collaborate with one another in pairs.
\end{enumerate}
Interesting studies could no doubt be made of these processes by combining
our network data with data on, for instance, institutional affiliations of
scientists.  Such studies are, however, perhaps better left to social
scientists and the social science journals.

The clustering coefficient of the MEDLINE database is worthy of brief
mention, since its value is far smaller than those for the other databases.
One possible explanation of this comes from the unusual social structure of
biomedical research, which, unlike the other sciences, has traditionally
been organized into laboratories, each with a ``principal investigator''
supervising a large number of postdocs, students, and technicians working
on different projects.  This organization produces a tree-like hierarchy of
collaborative ties with fewer interactions within levels of the tree than
between them.  A tree has no loops in it, and hence no triangles to
contribute to the clustering coefficient.  Although the biomedicine
hierarchy is certainly not a perfect tree, it may be sufficiently tree-like
for the difference to show up in the value of~$C$.  Another possible
explanation comes from the generous tradition of authorship in the
biomedical sciences.  It is common, for example, for a researcher to be
made a coauthor of a paper in return for synthesizing reagents used in an
experimental procedure.  Such a researcher will in many cases have a less
than average likelihood of developing new collaborations with their
collaborators' friends, and therefore of increasing the clustering
coefficient.

\section{Distances and centrality}
\label{centrality}
The basic statistics of the previous section are certainly of importance,
particularly for the construction of models such as those of
Refs.~\onlinecite{WS98,Kleinberg00,AJB99,NSW00}, but there is much more
that we can do with our collaboration networks.  In this section, we look
at some simple but useful measures of network structure, concentrating on
measures having to do with paths between vertices in the network.  In
Section~\ref{weighted} we discuss some shortcomings of these measures, and
construct some new and more complex measures which may better reflect true
collaboration patterns.

\subsection{Shortest paths}
\label{paths}
A fundamental concept in graph theory is the ``geodesic,'' or shortest path
of vertices and edges which links two given vertices.  There may not be a
unique geodesic between two vertices: there may be two or more shortest
paths, which may or may not share some vertices.  The geodesic(s) between
two vertices $i$ and $j$ can be calculated in time $\O(m)$, where $m$ is
the number of edges in the graph, using the following algorithm, which is a
modified form of the standard breadth-first search~\cite{Sedgewick88}.
\begin{enumerate}
\item Assign vertex $j$ distance zero, to indicate that it is zero steps
away from itself, and set $d\from0$.
\item For each vertex $k$ whose assigned distance is~$d$, follow each
attached edge to the vertex $l$ at its other end and if $l$ has not already
been assigned a distance, assign it distance $d+1$.  Declare $k$ to be a
predecessor of~$l$.
\item If $l$ has already been assigned distance~$d+1$, then there is no
need to do this again, but $k$ is still declared a predecessor of~$l$.
\item Set $d\from d+1$.
\item Repeat from step (2) until there are no unassigned sites left.
\end{enumerate}
Now the shortest path (if there is one) from $i$ to $j$ is the path you get
by stepping from $i$ to its predecessor, and then to the predecessor of
each successive site until $j$ is reached.  If a site has two or more
predecessors, then there are two or more shortest paths, each of which must
be followed separately if we wish to know all shortest paths from $i$
to~$j$.

\begin{figure}
\begin{center}
\psfig{figure=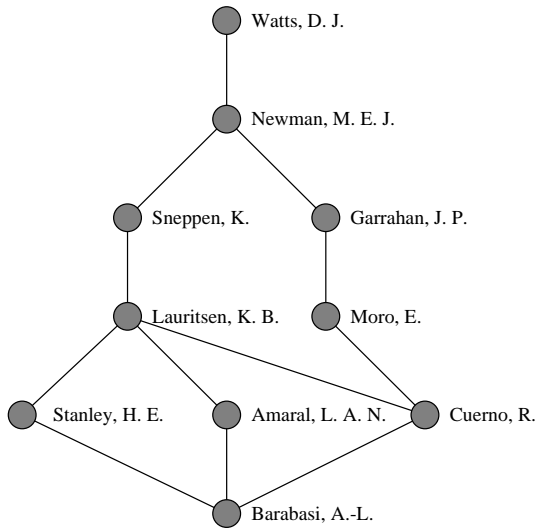,width=8cm}
\end{center}
\caption{The geodesics, or shortest paths, in the collaboration network of
physicists between Duncan Watts and Laszlo Barab\'asi.}
\label{djw2alb}
\end{figure}

In Fig.~\ref{djw2alb} we show the shortest paths of known collaborations
between two of the author's colleagues, Duncan Watts (Columbia) and Laszlo
Barab\'asi (Notre Dame), both of whom work on social networks of various
kinds.  It is interesting to note that, although the two scientists in
question are well acquainted both personally and with one another's work,
the shortest path between them does not run entirely through other
collaborations in the field.  (For example, the connection between the
present author and Juan Pedro Garrahan results from our coauthorship of a
paper on spin glasses.)  Although this may at first sight appear odd, it is
probably in fact a good sign.  It indicates that workers in the field come
from different scientific ``camps,'' rather than all descending
intellectually from a single group or institution.  This presumably
increases the likelihood that those workers will express independent
opinions on the open questions of the field, rather than merely spouting
slight variations on the same underlying doctrine.

A database which would allow one conveniently and quickly to extract
shortest paths between scientists in this way might have some practical
use.  Kautz~\etal~\cite{KSS97} have constructed a web-based system which
does just this for computer scientists, with the idea that such a system
might help to create new professional contacts by providing a ``referral
chain'' of intermediate scientists through whom contact may be established.

\subsection{Betweenness and funneling}
\label{betweenfun}
A quantity of interest in many social network studies is the
``betweenness'' of an actor~$i$, which is defined as the total number of
shortest paths between pairs of actors which pass through~$i$.  This
quantity is an indicator of who are the most influential people in the
network, the ones who control the flow of information between most others.
The vertices with high betweenness also result in the largest increase in
typical distance between others when they are removed~\cite{WF94}.

Naively, one might think that betweenness would take time of order
$\O(mn^2)$ to calculate for all vertices, since there are $\O(n^2)$
shortest paths to be considered, each of which takes time $\O(m)$ to
calculate, and standard network analysis packages such as
UCInet~\cite{UCInet} indeed use $\O(mn^2)$ algorithms at present.  However,
since breadth-first search algorithms can calculate $n$ shortest paths in
time $\O(m)$, it seems possible that one might be able to calculate
betweenness for all vertices in time $\O(mn)$.  Here we present a simple
new algorithm which performs this calculation.  Being enormously faster
than the standard packages, it makes possible exhaustive calculation of
betweenness on the very large graphs studied here.  The algorithm is as
follows.
\begin{enumerate}
\item The shortest paths from a vertex $i$ to every other vertex are
calculated using breadth-first search as described above, taking time
$\O(m)$.
\item A variable $b_k$, taking the initial value~1, is assigned to each
vertex~$k$.
\item Going through the vertices $k$ in order of their distance from $i$,
starting from the furthest, the value of $b_k$ is added to the
corresponding variable on the predecessor vertex of~$k$.  If $k$ has more
than one predecessor, then $b_k$ is divided equally between them.  This
means that if there are two shortest paths between a pair of vertices, the
vertices along those paths are given betweenness of $\half$ each.
\item When we have gone through all vertices in this fashion, the resulting
values of the variables $b_k$ represent the number of geodesic paths to
vertex $i$ which run through each vertex on the lattice, with the
end-points of each path being counted as part of the path.  To calculate
the betweenness for all paths, the $b_k$ are added to a running score
maintained for each site and the entire calculation is repeated for each of
the $n$ possible values of $i$.  The final running scores are precisely the
betweenness of each of the $n$ vertices.
\end{enumerate}

In column 3 of Table~\ref{top10} we show the ten highest betweennesses in
the {\tt astro-ph}, {\tt cond-mat}, and {\tt hep-th} subdivisions of the
Los Alamos Archive.  While we leave it to the knowledgeable reader to
decide whether the scientists named are indeed pivotal figures in their
respective fields, we do notice one interesting feature of the results.
The betweenness measure gives very clear winners in the competition: the
individuals with highest betweenness are well ahead of those with second
highest, who are in turn well ahead of those with third highest, and so on.
This same phenomenon has been noted in other social networks~\cite{WF94}.

Strogatz~\cite{StrogatzPC} has raised another interesting question about
social networks which we can address using our betweenness algorithm: are
all of your collaborators equally important for your connection to the rest
of the world, or do most paths from others to you pass through just a few
of your collaborators?  One could certainly imagine that the latter might
be true.  Collaboration with just one or two senior or famous members of
one's field could easily establish short paths to a large part of the
collaboration network, and all of those short paths would go through those
one or two members.  Strogatz calls this effect ``funneling.''  Since our
algorithm, as a part of its operation, calculates the vertices through
which each geodesic path to a specified actor $i$ passes, it is a trivial
modification to calculate also how many of those geodesic paths pass
through each of the immediate collaborators of that actor, and hence to use
it to look for funneling.

Our collaboration networks, it turns out, show strong funneling.  For most
people, their top few collaborators lie on most of the paths between
themselves and the rest of the network.  The rest of their collaborators,
no matter how numerous, account for only a small number of paths.
Consider, for example, the present author.  Out of the $44\,000$ scientists
in the giant component of the Los Alamos Archive collaboration network,
$31\,000$ paths from them to me, about 70\%, pass through just two of my
collaborators, Chris Henley and Juanpe Garrahan.  Another $13\,000$, most
of the remainder, pass through the next four collaborators.  The remaining
five account for a mere 1\% of the total.

\begin{figure}
\begin{center}
\psfig{figure=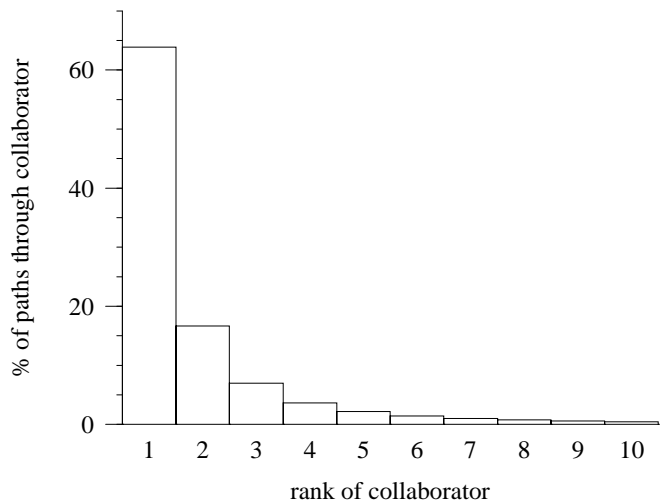,width=\columnwidth}
\end{center}
\caption{The average percentage of paths from other scientists to a given
scientist which pass through each collaborator of that scientist, ranked in
decreasing order.  The plot is for the Los Alamos Archive network, although
similar results are found for other networks.}
\label{funneling}
\end{figure}

\begin{figure*}[t]
\begin{center}
\psfig{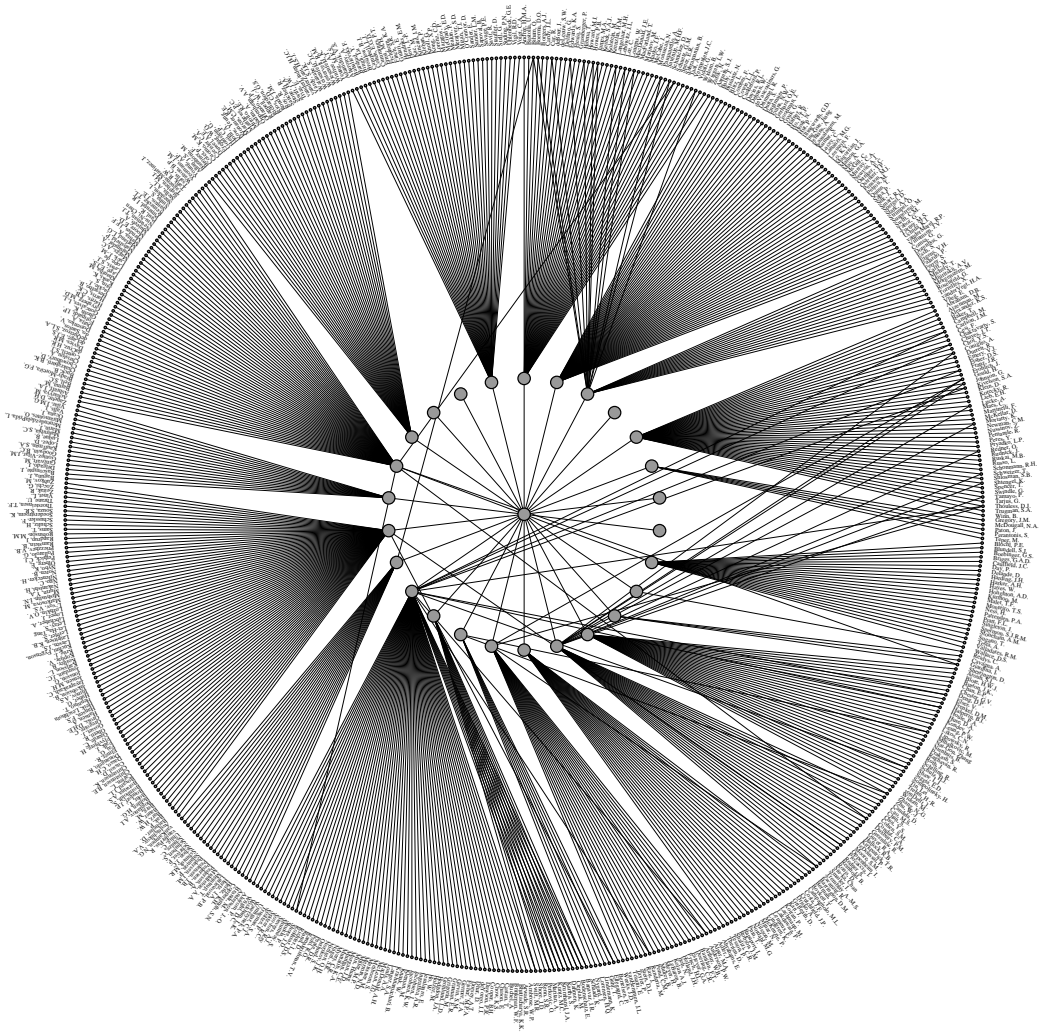}
\end{center}
\caption{The point in the center of the figure represents the author of the
paper you are reading, the first ring his collaborators, and the second
ring their collaborators.  Collaborative ties between members of the same
ring, of which there are many, have been omitted from the figure for
clarity.}
\label{wheel}
\end{figure*}

To give a more quantitative impression of the funneling effect, we show in
Fig.~\ref{funneling} the average fraction of paths that pass through the
top 10 collaborators of an author, averaged over all authors in the giant
component of the Los Alamos database.  The figure shows for example that on
average 64\% of one's shortest paths to other scientists pass through one's
top-ranked collaborator.  Another 17\% pass through the second-ranked one.
The top 10 shown in the figure account for 98\% of all paths.

That one's top few acquaintances account for most of one's shortest paths
to the rest of the world has been noted before in other contexts.  For
example, Stanley Milgram, in his famous ``small world''
experiment~\cite{Milgram67}, noted that most of the paths he found to a
particular target person in an acquaintance network went through just one
or two acquaintances of the target.  He called these acquaintances
``sociometric superstars.''

\subsection{Average distances}
\label{avdist}
Breadth-first search allows to us calculate exhaustively the lengths of the
shortest paths from every vertex on a graph to every other (if such a path
exists) in time~$\O(mn)$.  We have done this for each of the networks
studied here and averaged these distances to find the average distance
between any pair of (connected) authors in each of the subject fields
studied.  These figures are given in the penultimate row of
Table~\ref{summary}.  As the table shows, these figures are all quite
small: they vary from $4.0$ for SPIRES to $9.7$ for NCSTRL, although this
last figure may be artificially inflated by the poor coverage of this
database discussed in Section~\ref{gc}.  At any rate, all the figures are
very small compared to the number of vertices in the corresponding
databases.  This ``small world'' effect, first described by
Milgram~\cite{Milgram67}, is, like the existence of the giant component,
probably a good sign for science; it shows that scientific
information---discoveries, experimental results, theories---will not have
far to travel through the network of scientific acquaintance to reach the
ears of those who can benefit by them.  Even the {\em maximum\/} distances
between scientists in these networks, shown in the last row of the table,
are not very large, the longest path in any of the networks being just 31
steps long, again in the NCSTRL database, which may have poorer coverage
than the others.

The explanation of the small world effect is simple.  Consider
Fig.~\ref{wheel}, which shows all the collaborators of the present author
(in all subjects, not just physics), and all the collaborators of those
collaborators---all my first and second neighbors in the collaboration
network.  As the figure shows, I have 26 first neighbors, but 623 second
neighbors.  The ``radius'' of the whole network around me is reached when
the number of neighbors within that radius equals the number of scientists
in the giant component of the network, and if the increase in numbers of
neighbors with distance continues at the impressive rate shown in the
figure, it will not take many steps to reach this point.

\begin{figure}
\begin{center}
\psfig{figure=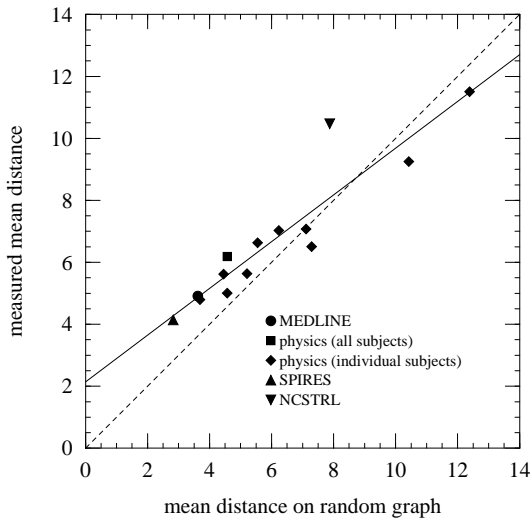,width=7cm}
\end{center}
\caption{Average distance between pairs of scientists in the various
networks, plotted against average distance on a random graph of the same
size and degree distribution.  The dotted line shows where the points would
fall if measured and predicted results agreed perfectly.  The solid line
is the best straight-line fit to the data.}
\label{z1z2size}
\end{figure}

This simple idea is borne out by theory.  In almost all networks, the
number of $k$th-nearest neighbors of a typical vertex increases
exponentially with $k$, and hence the average distance between pairs of
vertices $\ell$ scales logarithmically with $n$ the number of vertices.  In
a standard random graph~\cite{ER60,Bollobas85}, for instance, $\ell=\log
n/\log z$, where $z$ is the average degree of a vertex, the average number
of collaborators in our terminology.  In the more general class of random
graphs in which the distribution of vertex degrees is
arbitrary~\cite{MR9598}, rather than Poissonian as in the standard case,
the equivalent expression is~\cite{NSW00}
\begin{equation}
\ell = {\log(n/z_1)\over\log(z_2/z_1)} + 1,
\label{genell}
\end{equation}
where $z_1$ and $z_2$ are the average numbers of first and second neighbors
of a vertex.  It is in fact quite difficult to find a network which does
{\em not\/} show logarithmic behavior---such networks are a set of measure
zero in the limit of large~$n$.  Thus the presence of the small world
effect is hardly a surprise to anyone familiar with graph theory.  However,
it would be nice to demonstrate explicitly the presence of logarithmic
scaling in our networks.  Figure~\ref{z1z2size} does this in a crude
fashion.  In this figure we have plotted the measured value of $\ell$, as
given in Table~\ref{summary}, against the value given by Eq.~\eref{genell}
for each of our four databases, along with separate points for nine of the
subject-specific subdivisions of the Los Alamos Archive.  As the figure
shows, the correlation between measured and predicted values is quite good.
A straight-line fit has $R^2=0.86$, rising to $R^2=0.95$ if the NCSTRL
database, with its incomplete coverage, is excluded (the downward-pointing
triangle in the figure).

Figure~\ref{z1z2size} needs to be taken with a pinch of salt.  Its
construction implicitly assumes that the different networks are
statistically similar to one another and to random graphs with the same
distributions of vertex degree, an assumption which is almost certainly not
correct.  In practice, however, the measured value of $\ell$ seems to
follow Eq.~\eref{genell} quite closely.  Turning this observation around,
our results also imply that it is possible to make a good prediction of the
typical vertex--vertex distance in a network by making only local
measurements of the average numbers of neighbors that vertices have.  If
this result extends beyond coauthorship networks to other social networks,
it could be of some importance for empirical work, where the ability to
calculate global properties of a network by making only local measurements
could save large amounts of effort.


We can also trivially use our breadth-first search algorithm to calculate
the average distance from a single vertex to all other vertices in the
giant component.  This average is essentially the same as the quantity
known as ``closeness'' to social network analysts.  Like betweenness it is
also a measure, in some sense, of the centrality of a vertex---authors with
low values of this average will, it is assumed, be the first to learn new
information, and information originating with them will reach others
quicker than information originating with other sources.  Average distance
is thus a measure of centrality of an actor in terms of their access to
information, unlike betweenness which is a measure of an actor's control
over information flowing between others.

\begin{figure}
\begin{center}
\psfig{figure=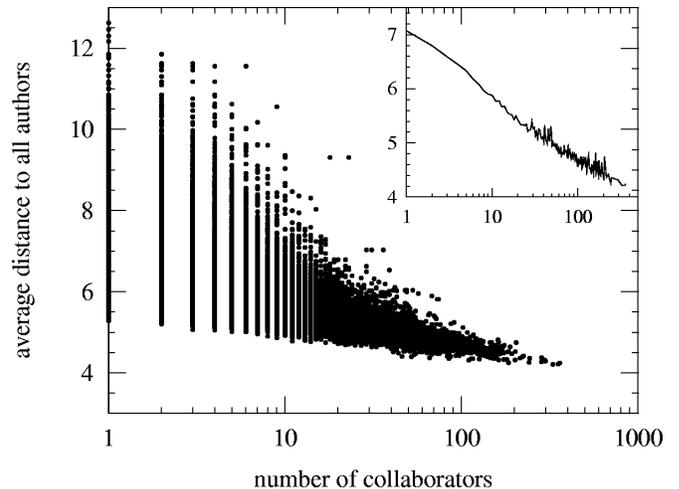,width=\columnwidth}
\end{center}
\caption{Scatter plot of the mean distance from each physicist in the giant
component of the Los Alamos Archive data to all others as a function of
number of collaborators.  Inset: the same data averaged vertically over all
authors having the same number of collaborators.}
\label{closeness}
\end{figure}

Calculating average distance for many networks returns results which look
sensible to the observer.  Calculations for the network of collaborations
between movie actors, for instance, gives small average distances for
actors who are famous---ones many of us will have heard of.  Interestingly,
however, performing the same calculation for our scientific collaboration
networks does not return sensible results.  For example, one finds that the
people at the top of the list are always experimentalists.  This, you might
think, is not such a bad thing: perhaps the experimentalists are better
connected people?  In a sense, in fact, it turns out that they are.  In
Fig.~\ref{closeness} we show the average distance from scientists in the
Los Alamos Archive to all others in the giant component as a function of
their number of collaborators.  As the figure shows, there is a trend
towards shorter average distance as the number of collaborators becomes
large.  This trend is clearer still in the inset, where we show the same
data averaged over all authors who have the same number of collaborators.
Since experimentalists work in large groups, it is not surprising to learn
that they tend to have shorter average distances to other scientists.

But this brings up an interesting question, one that we touched upon in
Section~\ref{collabnets}: while most pairs of people who have written a
paper together will know one another reasonably well, there are exceptions.
On a high-energy physics paper with 1000 coauthors, for instance, it is
unlikely that every one of the $499\,500$ possible acquaintanceships
between pairs of those authors will actually be realized.  Our closeness
measure does not take into account the tendency for collaborators in large
groups not to know one another, or to know one another less well.  In the
next section we study a more sophisticated form of collaboration network
which does do this.

\section{Weighted collaboration networks}
\label{weighted}
There is more information present in the databases used here than in the
simple networks we have constructed from them, which tell us only whether
scientists have collaborated or not~\cite{note4}.  In particular, we know
on how many papers each pair of scientists has collaborated during the
period of the study, and how many other coauthors they had on each of those
papers.  We can use this information to make an estimate of the strength of
collaborative ties.

First of all, it is probably the case, as we pointed out at the end of the
previous section, that two scientists whose names appear on a paper
together with many other coauthors know one another less well on average
than two who were the sole authors of a paper.  The extreme case of a very
large collaboration which we discussed illustrates this point forcefully,
but it probably applies to smaller collaborations too.  Even on a paper
with four or five authors, the authors probably know one another less well
on average than authors from a smaller collaboration.  To account for this
effect, we weight collaborative ties inversely according to the number of
coauthors as follows.  Suppose a scientist collaborates on the writing of a
paper which has $n$ authors in total, i.e.,~he or she has $n-1$ coauthors
on that paper.  Then we assume that he or she is acquainted with each
coauthor $1/(n-1)$ times as well, on average, as if there were only one
coauthor.  One can imagine this as meaning that the scientist divides his
or her time equally between the $n-1$ coauthors.  This is obviously only a
rough approximation: in reality a scientist spends more time with some
coauthors than with others.  However, in the absence of other data, it is
the obvious first approximation to make~\cite{note5}.

\begin{figure}
\begin{center}
\psfig{figure=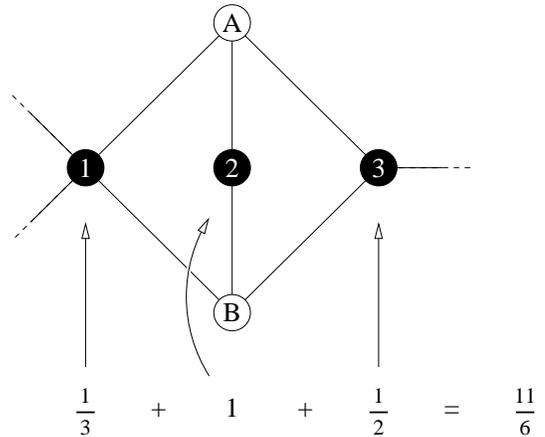,width=7cm}
\end{center}
\caption{Authors A and B have coauthored three papers together, labeled 1,
2, and 3, which had respectively four, two, and three authors.  The tie
between A and B accordingly accrues weight $\frac13$, $1$, and $\frac12$
from the three papers, for a total weight of $\frac{11}{6}$.}
\label{weightcalc}
\end{figure}

Second, authors who have written many papers together will, we assume, know
one another better on average than those who have written few papers
together.  To account for this, we add together the strengths of the ties
derived from each of the papers written by a particular pair of
individuals~\cite{ML00}.  Thus, if $\delta_i^k$ is one if scientist $i$ was
a coauthor of paper $k$ and zero otherwise, then our weight $w_{ij}$
representing the strength of the collaboration (if any) between scientists
$i$ and $j$ is
\begin{equation}
w_{ij} = \sum_k {\delta_i^k\delta_j^k\over n_k-1},
\label{wij}
\end{equation}
where $n_k$ is the number of coauthors of paper~$k$ and we explicitly
exclude from our sums all single-author papers.  (They do not contribute to
the coauthorship network, and their inclusion in Eq.~\eref{wij} would make
$w_{ij}$ ill-defined.)  We illustrate this measure for a simple example in
Fig.~\ref{weightcalc}.

Note that the equivalent of vertex degree for our weighted
network---i.e.,~the sum of the weights for each of an individual's
collaborations---is now just equal to the number of papers they have
coauthored with others:
\begin{equation}
\sum_{j(\ne i)} w_{ij} = \sum_k\sum_{j(\ne i)}
                         {\delta_i^k\delta_j^k\over n_k-1}
                       = \sum_k \delta_i^k.
\end{equation}

\begin{figure}
\begin{center}
\psfig{figure=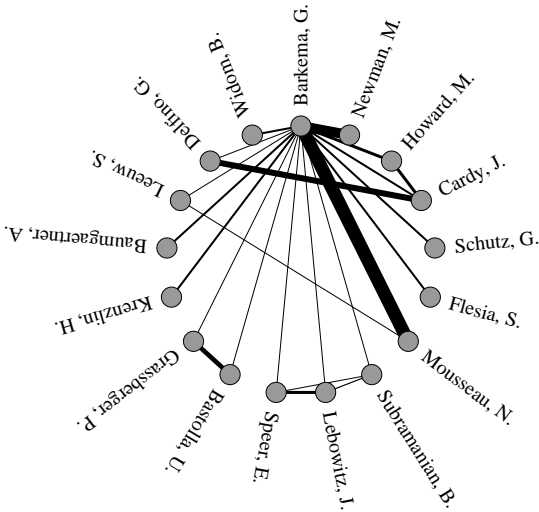,width=\columnwidth}
\end{center}
\caption{Gerard Barkema and his collaborators, with lines representing
collaborations whose thickness is proportional to our estimate,
Eq.~\eref{wij}, of the strength of the corresponding tie.}
\label{barkema}
\end{figure}

In Fig.~\ref{barkema} we show as an example collaborations between Gerard
Barkema (one of the present author's frequent collaborators), and all of
his collaborators in the Los Alamos Archive for the past five years.  Lines
between points represent collaborations, with their thickness proportional
to the weights $w_{ij}$ of Eq.~\eref{wij}.  As the figure shows, Barkema
has collaborated closely with myself and with Normand Mousseau, and less
closely with a number of others.  Also, two of his collaborators, John
Cardy and Gesualdo Delfino have collaborated quite closely with one
another.


In the last column of Table~\ref{top10} we show the pairs of collaborators
who have the strongest collaborative ties in three subdivisions of the Los
Alamos Archive.

We have used our weighted collaboration graphs to calculate distances
between scientists.  In this simple calculation we assumed that the
distance between authors is just the inverse of the weight of their
collaborative tie.  Thus if one pair of authors know one another twice as
well as another pair, the distance between them is half as great.
Calculating minimum distances between vertices on a weighted graph such as
this cannot be done using the breadth-first search algorithm of
Section~\ref{paths}, since the shortest weighted path may not be the
shortest in terms of number of steps on the unweighted network.  Instead we
use Dijkstra's algorithm~\cite{AMO93}, which calculates all distances from
a given starting vertex $i$ as follows.
\begin{enumerate}
\item Distances from vertex $i$ are stored for each vertex and each is
labeled ``exact,'' meaning we have calculated that distance exactly, or
``estimated,'' meaning we have made an estimate of the distance, but that
estimate may be wrong.  We start by assigning an estimated distance of
$\infty$ to all vertices except vertex $i$ to which we assign an estimated
distance of zero.  (We know the latter to be exactly correct, but for the
moment we consider it merely ``estimated.'')
\item From the set of vertices whose distances from $i$ are currently
marked ``estimated,'' choose the one with the lowest estimated distance,
and mark this ``exact.''
\item Calculate the distance from that vertex to each of its immediate
neighbors in the network by adding to its distance the length of the edges
leading to those neighbors.  Any of these distances which is shorter than a
current estimated distance for the same vertex supersedes that current
value and becomes the new estimated distance for the vertex.
\item Repeat from step (2), until no ``estimated'' vertices remain.
\end{enumerate}
A naive implementation of this algorithm takes time $\O(mn)$ to calculate
distances from a single vertex to all others, or $\O(mn^2)$ to calculate
all pairwise distances.  One of the factors of $n$, however, arises because
it takes time $\O(n)$ to search through the vertices to find the one with
the smallest estimated distance.  This operation can be improved by storing
the estimated distances in a binary heap (a partially ordered binary tree
with its smallest entry at its root).  We can find the smallest distance in
such a heap in time $\O(1)$, and add and remove entries in time $\O(\log
n)$.  This speeds up the operation of the algorithm to $\O(mn\log n)$,
making the calculation feasible for the large networks studied here.

\begin{table}
\setlength{\tabcolsep}{4pt}
\begin{center}
\begin{tabular}{lrl|c|c}
 & rank & name & co-workers & papers \\
\hline
 {\tt astro-ph}: & 1  & Rees, M. J.        &  31 &  36 \\
                 & 2  & Miralda-Escude, J. &  36 &  34 \\
                 & 3  & Fabian, A. C.      & 156 & 112 \\
                 & 4  & Waxman, E.         &  15 &  30 \\
                 & 5  & Celotti, A.        & 119 &  45 \\
                 & 6  & Narayan, R.        &  65 &  58 \\
                 & 7  & Loeb, A.           &  33 &  64 \\
                 & 8  & Reynolds, C. S.    &  45 &  38 \\
                 & 9  & Hernquist, L.      &  62 &  80 \\
                 & 10 & Gould, A.          &  76 &  79 \\
\hline
 {\tt cond-mat}: & 1  & Fisher, M. P. A.   &  21 &  35 \\
                 & 2  & Balents, L.        &  24 &  29 \\
                 & 3  & MacDonald, A. H.   &  64 &  70 \\
                 & 4  & Senthil, T.        &   9 &  13 \\
                 & 5  & Das Sarma, S.      &  51 &  75 \\
                 & 6  & Millis, A. J.      &  43 &  37 \\
                 & 7  & Ioffe, L. B.       &  16 &  27 \\
                 & 8  & Sachdev, S.        &  28 &  44 \\
                 & 9  & Lee, P. A.         &  24 &  34 \\
                 & 10 & Jungwirth, T.      &  27 &  17 \\
\hline
 {\tt hep-th}:   & 1  & Cvetic, M.         &  33 &  69 \\
                 & 2  & Behrndt, K.        &  22 &  41 \\
                 & 3  & Tseytlin, A. A.    &  22 &  65 \\
                 & 4  & Bergshoeff, E.     &  21 &  39 \\
                 & 5  & Youm, D.           &   3 &  30 \\
                 & 6  & Lu, H.             &  34 &  73 \\
                 & 7  & Klebanov, I. R.    &  29 &  47 \\
                 & 8  & Townsend, P. K.    &  31 &  54 \\
                 & 9  & Pope, C. N.        &  33 &  72 \\
                 & 10 & Larsen, F.         &  11 &  27 \\
\end{tabular}
\end{center}
\caption{The ten best connected individuals in three of the communities
studied here, calculated using the weighted distance measure described in
the text.}
\label{winners}
\end{table}

It is in theory possible to generalize any of the calculations of
Section~\ref{centrality} to the weighted collaboration graph using this
algorithm and variations on it.  For example, we can find shortest paths
between specified pairs of scientists, as a way of establishing referrals.
We can calculate the weighted equivalent of betweenness by a simple
adaption of our fast algorithm of Section~\ref{betweenfun}---we use
Dijkstra's algorithm to establish the hierarchy of predecessors of vertices
and then count paths through vertices exactly as before.  We can also study
the weighted version of the ``funneling'' effect using the same algorithm.
Here we carry out just one calculation explicitly to demonstrate the idea;
we calculate the weighted version of the distance centrality measure of
Section~\ref{avdist}, i.e.,~the average weighted distance from a vertex to
all others.  In Table~\ref{winners} we show the winners in this particular
popularity contest, along with their numbers of collaborators and papers in
the database.  Many of the scientists who score highly here do indeed
appear to be well connected individuals.  For example, number~1 best
connected astrophysicist, Martin Rees, is the Astronomer Royal of Great
Britain.  (On being informed of this latest honor, Prof.~Rees is reported
as replying, ``I'm certainly relieved not to be the most disconnected
astrophysicist''~\cite{Muir00}.)

What is interesting to note however (apart from nonchalantly checking to
see if one has made it into the top~10) is that sheer number of
collaborators is no longer a necessary prerequisite for being
well-connected in this sense (although some of the scientists listed do
have a large number of collaborators).  The case of D.~Youm is particularly
startling, since Youm has only three collaborators listed in the database,
but nonetheless is fifth best connected high-energy theorist (out of eight
thousand), because those three collaborators are themselves very well
connected, and because their ties to Youm are very strong.
Experimentalists no longer dominate the field, although the well-connected
amongst them still score highly.

Note that the number of papers for each of the well-connected scientists
listed is high.  Having written a large number of papers is, as it rightly
should be, always a good way of becoming well connected.  Whether you write
many papers with many different authors, or many with a few, writing many
papers will put you in touch with your peers.

\section{Conclusions}
\label{concs}
In this paper we have studied social networks of scientists in which the
actors are authors of scientific papers, and a tie between two authors
represents coauthorship of one or more papers.  Drawing on the lists of
authors in four databases of papers in physics, biomedical research, and
computer science, we have constructed explicit networks for papers
appearing between the beginning of 1995 and the end of 1999.  We have
cataloged a large number of basic statistics for our networks, including
typical numbers of papers per author, authors per paper, and numbers of
collaborators per author in the various fields.  We also note that the
distributions of these quantities roughly follow a power-law form, although
there are some deviations which may be due to the finite time window used
for the study.

We have also looked at a variety of non-local properties of our networks.
We find that typical distances between pairs of authors through the
networks are small---the networks form a ``small world'' in the sense
discussed by Milgram---and that they scale logarithmically with total
number of authors in a network, in reasonable agreement with the
predictions of random graph models.  We have introduced a new algorithm for
counting the number of shortest paths between vertices on a graph which
pass through each other vertex, which is one order of system size faster
than previous algorithms, and used this to calculate the so-called
``betweenness'' measure of centrality on our graphs.  We also show that for
most authors the bulk of the paths between them and other scientists in the
network go through just one or two or their collaborators, an effect which
Strogatz has dubbed ``funneling.''

We have suggested a measure of the closeness of collaborative ties which
takes account of the number of papers a given pair of scientists have
written together, as well as the number of other coauthors with whom they
wrote them.  Using this measure we have added weightings to our
collaboration networks to represent closeness and used the resulting
networks to find those scientists who have the shortest average distance to
others.  Generalization of the betweenness and funneling calculations to
these weighted networks is also straightforward.

The calculations presented in this paper inevitably represent only a small
part of the investigations which could be conducted using large network
datasets such as these.  Indeed, to a large extent, the purpose of this
paper is simply to alert other researchers to the presence of a valuable
source of network data in bibliometric databases, and to provide a first
demonstration of the kind of results which can be derived from them.  We
hope, given the high current level of interest in network phenomena, that
others will find many further uses for collaboration network data.

\section*{Acknowledgements}
The author would particularly like to thank Paul Ginsparg for his
invaluable help in obtaining the data used for this study.  The data used
were generously made available by Oleg Khovayko, David Lipman, and Grigoriy
Starchenko (MEDLINE), Paul Ginsparg and Geoffrey West (Los Alamos E-print
Archive), Heath O'Connell (SPIRES), and Carl Lagoze (NCSTRL).  The Los
Alamos E-print Archive is funded by the NSF under grant number
PHY--9413208.  NCSTRL is funded through the DARPA/CNRI test suites program
under DARPA grant N66001--98--1--8908.

The author also would like to thank Steve Strogatz for suggesting the
``funneling effect'' calculation of Section~\ref{betweenfun}, Laszlo
Barab\'asi for making available an early version of Ref.~\cite{Barabasi00},
and Dave Alderson, Laszlo Barab\'asi, Sankar Das Sarma, Paul Ginsparg, Rick
Grannis, Jon Kleinberg, Laura Land\-weber, Vito Latora, Hazel Muir, Sid
Redner, Ronald Rousseau, Steve Strogatz, Duncan Watts, and Doug White for
many useful comments and suggestions.  This work was funded in part by a
grant from Intel Corporation to the Santa Fe Institute's Network Dynamics
Program.

\medbreak
{\em Note added:\/} After this work was completed the author learned of a
recent preprint by Brandes~\cite{Brandes00} in which an algorithm for
calculating betweenness similar to ours is described.  The author is
grateful to Rick Grannis for bringing this to his attention.


\begin{references}
\bibitem{note4}
A complete description of a collaboration network, or indeed any
affiliation network, requires us to construct a bipartite graph or
hypergraph of actors and the groups to which they belong~\cite{WF94,NSW00}.
For our present purposes however, such detailed representations are not
necessary.
\bibitem{note5}
One can imagine using a measure of connectedness which weights authors more
or less heavily depending on the order in which their names appear on a
publication.  We have not adopted this approach here however, since it will
probably discriminate against those authors with names falling towards the
end of the alphabet, who tend to find themselves at the ends of purely
alphabetical author lists.
\end{references}
\end{document}